%% file: Lattice2017_232_VADACCHINO.tex
\documentclass[epj]{webofc}
\usepackage[utf8]{inputenc}
\usepackage[varg]{txfonts}   % Web of Conferences font
\usepackage{booktabs}
\usepackage{xcolor}
\definecolor{darkred}{rgb}{0.4,0.0,0.0}
\definecolor{darkgreen}{rgb}{0.0,0.4,0.0}
\definecolor{darkblue}{rgb}{0.0,0.0,0.4}
\usepackage[bookmarks,linktocpage,colorlinks,
    linkcolor = darkred,
    urlcolor  = darkblue,
    citecolor = darkgreen]{hyperref}
\usepackage{amsmath}
\usepackage[labelfont=bf,font=small]{caption}
\usepackage{subcaption}

\wocname{EPJ Web of Conferences}
\woctitle{Lattice2017}
\newcommand{\Tr}{\mathrm{Tr}}
\newcommand{\Sp}[1]{\ensuremath{\mathrm{Sp}(#1)}}
\newcommand{\SU}[1]{\ensuremath{\mathrm{SU}(#1)}}
\DeclareMathOperator{\re}{Re}
\DeclareMathOperator{\tr}{tr}
\begin{document}
%%%%%%%%%%%%%%%%%%%%%%%%%%%%%%%%%%%%%%%%%%%%%%%%%%%%%%%%%%%%%%%%%%%%%%%%%%%%%
%
\selectlanguage{english}
%----------------------------------------------------------------------------
\title{%
Higgs compositeness in $\Sp{2N}$ gauge theories --- \\The pure gauge model
}
%%----------------------------------------------------------------------------
\author{%
\firstname{Ed} \lastname{Bennett}\inst{1}\fnsep\thanks{We acknowledge the support of the Supercomputing Wales project, which is part-funded by the European Regional Development Fund (ERDF) via Welsh Government.} \and
\firstname{Deog Ki} \lastname{Hong}\inst{2}\fnsep\thanks{Supported in part by Korea Research Fellowship program funded by the Ministry of Science, ICT and Future Planning through the National Research Foundation of Korea (2016H1D3A1909283) and under the framework of international cooperation program (NRF-2016K2A9A1A01952069).} \and
\firstname{Jong-Wan}  \lastname{Lee}\inst{3}\fnsep\inst{2}\fnsep\footnotemark[2] \and
\firstname{C.-J.~David} \lastname{Lin}\inst{4}\fnsep\thanks{Supported by Taiwanese MoST grant 105-2628-M-009-003-MY4.} \and \\
\firstname{Biagio} \lastname{Lucini}\inst{5}\fnsep\thanks{Supported in part by the Royal Society and the Wolfson Foundation.}\fnsep\thanks{Supported in part by the STFC Consolidated Grants ST/L000369/1 and ST/P00055X/1.} \and
\firstname{Maurizio} \lastname{Piai}\inst{6}\fnsep\footnotemark[5] \and
\firstname{Davide} \lastname{Vadacchino}\inst{1}\fnsep\thanks{Speaker, \email{d.vadacchino@swansea.ac.uk}.}\fnsep\footnotemark[1]\fnsep\footnotemark[5]
% etc.
}
%\author{%
%\firstname{Ed} \lastname{Bennett}\inst{1}\fnsep\thanks{Funded by the Supercomputing Wales project, which is part-funded by the European Regional Development Fund (ERDF) via Welsh Government.} \and
%\firstname{Deog Ki} \lastname{Hong}\inst{2}\fnsep\footnotemark[2] \and
%\firstname{Jong-Wan}  \lastname{Lee}\inst{3}\fnsep\thanks{The work of DKH and JWL was supported in part by Korea Research Fellowship program funded by the Ministry of Science, ICT and Future Planning through the National Research Foundation of Korea (2016H1D3A1909283) and under the framework of international cooperation program (NRF-2016K2A9A1A01952069).} \and
%\firstname{C.-J.~David} \lastname{Lin}\inst{4}\thanks{Supported by Taiwanese MoST grant 105-2628-M-009-003-MY4} \and
%\firstname{Biagio} \lastname{Lucini}\inst{5}\fnsep\thanks{Supported in part by the Royal Society and the Wolfson Foundation.}\fnsep\thanks{Supported in part by the STFC Consolidated Grants ST/L000369/1 and ST/P00055X/1.} \and
%\firstname{Maurizio} \lastname{Piai}\inst{6}\fnsep\footnotemark[3] \and
%\firstname{Davide} \lastname{Vadacchino}\inst{1}\fnsep\thanks{Speaker, \email{d.vadacchino@swansea.ac.uk}}\fnsep\footnotemark[1]\fnsep\footnotemark[3]
%% etc.
%}
%----------------------------------------------------------------------------
\institute{%
Swansea Academy of Advanced Computing, Swansea University, Singleton Park, Swansea, SA2 8PP, UK
\and
Department of Physics, Pusan National University, Busan 46241, Korea
\and
Extreme Physics Institute, Pusan National University, Busan 46241, Korea
\and
Institute of Physics, National Chiao-Tung University, Hsinchu 30010, Taiwan
\and
Department of Mathematics, Swansea University, Singleton Park, Swansea, SA2 8PP, UK
\and
Department of Physics, Swansea University, Singleton Park, Swansea, SA2 8PP, UK
}
\abstract{%
As a first step in the study of $\Sp{2N}$ composite Higgs models, we obtained a set of novel numerical 
results for the pure gauge $\Sp{4}$ lattice theory in 3+1 space-time dimensions. Results for the continuum 
extrapolations of the string tension and the glueball mass spectrum are presented 
and their values are compared with the same quantities in neighbouring SU(N) models.
}
%----------------------------------------------------------------------------
\maketitle
%----------------------------------------------------------------------------
\section{Introduction}\label{intro}

The $\Sp{2N}$ class of gauge theories naturally arises in the context of Higgs compositeness whenever the symmetry group of the new strongly coupled sector is pseudoreal. As explained in~\cite{thiscontrib182} in more detail, it is then interesting to perform a non-perturbative study on this class of models using their lattice regularization. In this contribution we focus on the glueball spectrum and string tension computation in the $\Sp{4}$ lattice regularized gauge theory. The first part is devoted to the implementation of the heat bath (HB) algorithm used to simulate the $\Sp{2N}$ gauge theory. In the second part, we focus on variational methods and improved operators employed to obtain the glueball spectrum and the string tension of the theory. The last part is a discussion of the results obtained for $N=2$, i.e. $\Sp{4}$.  

\section{The heat bath algorithm for $\Sp{2N}$}\label{HBsp2n}

The $\Sp{2N}$ lattice gauge theory is defined on a 4 dimensional euclidean hypercubic lattice of spacing $a$ by the Wilson action
\begin{equation}
	S = \beta\sum_{x,\mu>\nu}\left( 1 - \frac{1}{2N} \re \tr U_{\mu\nu}\right)
\end{equation}
where $\beta=4N/g^2$ and
\begin{equation}
	U_{\mu\nu}(x) = U_\mu(x)~ U_\nu(x+\hat{\mu})~ U_\mu^\dag(x+\hat{\nu})~ U_\nu^\dag(x)~,\qquad U_\mu(x)\in \Sp{2N}
\end{equation}

An update algorithm for the above theory can be realized following the well known Cabibbo-Marinari approach~\cite{Cabibbo:1982zn,Holland:2003kg}. The crucial observation is that since $\Sp{2N}$ is a subgroup of $\SU{2N}$, an ergodic update algorithm can be obtained for the former group by restricting the set of usually updated $\SU{2}$ subgroups of the latter. Starting from the unit matrix (cold start) or by a randomly chosen $\Sp{2N}$ matrix (hot start), and updating the lattice links $U_\mu(x)$ according to the above scheme, one samples the configuration space in the desired way. 

\begin{figure}[thb] % no figure before 1st section
	\begin{subfigure}{0.5\textwidth}
	\centering
  	\includegraphics[width=0.9\textwidth,clip]{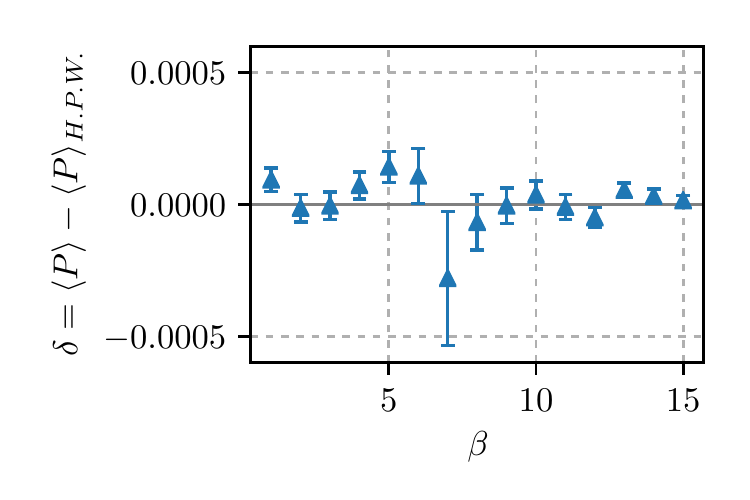}
\end{subfigure}
\begin{subfigure}{0.5\textwidth}
  	\centering
  	\includegraphics[width=0.9\textwidth,clip]{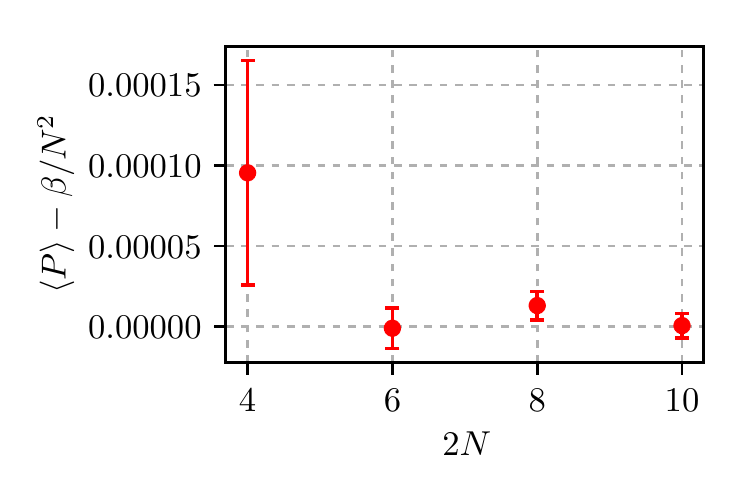}
\end{subfigure}
  	\caption{To check for the correctness of our update algorithm, we compared our results for the expectation value of the action per plaquette at various values of $\beta$ both with those present in the literature and with analytical calculations. The deviations between the estimates obtained in the present work and those of ~\cite{Holland:2003kg} (left), and with the leading order strong coupling expectation (right).}\label{fig:plaqvsbeta}% Give a unique label
\end{figure}

To check the correctness of the above procedure, we measured the vacuum expectation value of the action per plaquette $P =\tfrac{1}{6V} S $ for several values of $\beta$ and at $4$ values of $N$ starting from both hot and cold configurations, and compared our estimates with the strong and weak coupling expansions and with known numerical results~\cite{Holland:2003kg}. The deviations with respect to the latter at $N=2$ is shown on the left hand side fig.~\ref{fig:plaqvsbeta} while on the right hand side of fig.~\ref{fig:plaqvsbeta} we show the deviation with respect to the leading order strong coupling prediction at several values of $N$. These results suggest that we are able to simulate the lattice regularized $\Sp{2N}$ gauge theory at any value of $N$ and of the coupling $\beta$.

\section{Methods for Glueballs spectrum and string tension calculations}

In the confining phase of the $\Sp{2N}$ gauge theory we expect two kinds of color-singlet states to propagate: bound states of gluons called glueballs and, in the presence of static sources or in compactified space-times, closed fluxtubes. The quantity we will estimate is the ratio $\frac{m_G}{\sqrt{\sigma}}$, where $m_G$ is the mass of the glueball in a particular symmetry channel and $\sigma$ the energy per unit length of a fluxtube, i.e. the string tension. In this section, we explain how both of these quantities can be estimated on the lattice.

In the continuum, states of the gauge field can be labelled by their energy value, by the values of their conserved charges with respect to the euclidean Poincar\'e group and, in the case of fluxtubes, by their length. For glueballs, the relevant quantum numbers are the angular momentum $J$ and parity $P$, while for fluxtubes, we use, in addition, their length $L$, the flux $q$ they carry and parities $P_l, P_t$, respectively in the plane of the fluxtube and with respect to its axis. Therefore, glueballs are labelled by $(J,P)$ and fluxtubes by $(L,J,q,P_l,P_t)$. Since for the latter we only deal with the case $q=1$, $J=0$ and $P_l=P_t=1$, we will omit the corresponding labels. Moreover, since $\Sp{2N}$ has only pseudo real representations, the charge conjugation quantum number has always value $+1$ and will neglected below. 

The lattice formulation of the gauge theory breaks its Poincar\'e symmetry down to its discrete icosahedral subgroup. States created on the lattice by gauge invariant products of elementary link operators $U_\mu(\vec{x},t)$ along closed loops $\Omega$ will be labelled by the irreducible representation of the icosahedral group in which they transform. The latter can be labelled as $A_1, A_2, T_1, T_2, E$. A further distinction has to be made between contractible and non-contractible loop operators, generating, respectively, glueball and fluxtube states. Note that because of center symmetry, these have no overlap between each other and can be studied separately.

Now we will briefly review how to obtain the mass $m_C$ of a generic state of quantum numbers $C$. As said above, $C = R^{\pm}$ for glueballs, while $C=L$ for fluxtubes. Denoting by
\begin{equation}\label{eq:glue-op}
	\phi_C(\vec{x},t) = \Tr \prod_\Omega U_\mu(\vec{x},t)~,
\end{equation}
a generic gauge-invariant operator with quantum numbers $C$, the mass $m_C$ of the corresponding state can be obtained from the asymptotic behaviour of its zero-momentum-projected euclidean time correlator,
\begin{equation}\label{eq:glue-corr-1}
%\langle phi_C(t) phi_C^\dag(t)\rangle = \sum_n \langle vac | phi_C(t) | J,P,n\rangle \langle J,P,n | phi_C^\dag(0) |vac\rangle
\Gamma_C(t) =	\frac{\langle \phi_C(t) \phi_C^\dag(0)\rangle}{\langle \phi_C(0) \phi_C^\dag(0)\rangle} = \sum_n \frac{|\langle C,n | \phi_C^\dag(0) |vac\rangle|^2}{\sum_m |\langle C,m | \phi_C^\dag(0) |vac\rangle|^2} e^{-m_{C,n} t}
\end{equation} 
where $n$ labels the excitation number and $\phi_C(t) = \sum_{\vec{x}} \phi(\vec{x},t)$.
%\begin{equation}
%\phi_C(t) = \sum_{\vec{x}} \phi(\vec{x},t)~.
%\end{equation}
The above correlator probes the propagation of a infinite tower of states of increasing energy with quantum numbers $C$. The leading behaviour for $t\to\infty$ singles out the lowest among the $m_{C,n}$, that we denote simply by $m_C$ from now on,
\begin{equation}
m_{C} = - \lim_{t\to\infty} \frac{\log \Gamma_C(t)}{t}~.
\end{equation}

Direct estimates of $m_C$ can be obtained by fitting measurements of the correlator in eq.~(\ref{eq:glue-corr-1}) on a suitable range of $t$ with an exponentially decaying function. In this process one has to face two severe difficulties, however. Not only does the signal-to-noise ratio exponentially decay at large $t$, but, as we increase $\beta$ to flow towards the continuum limit, the amplitude of the exponential decay rapidly goes to $0$, turning the \emph{direct} determination of $m_C$ into a very challenging task. These are by now well known problems related to our choice of interpolating operators, eq.~(\ref{eq:glue-op}), and to their behaviour in the continuum limit, for which a solution has been found in the form of variational calculus with linear combinations of improved operators. Improved operators are the result of two iterative operations, smearing and blocking, that are designed in order to interpolate the physical size of lattice states as its spacing is brought to $0$. Smearing consists in summing the staples around a link to the link itself, as follows
\begin{align}
	\tilde{U}^{s+1}_i(n) &= U^s_i(n) + p_a \sum_{j\neq i} U^s_j(n) U^s_i(n+j) U^{s\dag}_j(n+i) + p_a \sum_{j\neq i} U^{s\dag}_j(n-j) U^{s}_i(n-j) U^{s}_j(n-j+i)\\
	U^{s=0}_i(n) &= U_i(n)
\end{align}
where $p_a$ is a free parameter that determines how many smearing steps are necessary to reach a particular scale. Blocking consists in replacing the original elementary links with \emph{superlinks} that join lattice sites that are $2^b$ spacings apart, where $b$ is the number of blocking iterations, as described by
\begin{align}
	\tilde{U}^{b+1}_i(n) &= U^b_i(n)U^b_i(n+2^bi) + p_b \sum_{j\neq i} U^b_j(n) U^b_i(n+2^bj) U^b_i(n+2^bj+2^bi) U^{b\dag}_j(n+2^bi)\\
	&+ p_b \sum_{j\neq i} U^{b\dag}_j(n-2^bj) U^{s}_i(n-2^bj)U^{s}_i(n-2^bj+2^bi) U^{b}_j(n-2^bj+2^bi)\\
	U^{b=0}_i(n) &= U_i(n)
\end{align}

While blocking allows to reach the physical size of the glueball with less steps, at the physical scale smearing provides a better resolution. An iterative combination of $n=1,2$ smearing steps with a blocking step has shown to be an efficient strategy~\cite{Lucini:2010nv}, and we have employed this procedure in this work, with $(p_a=0.4,~p_b=0.16)$. Note that the matrices $\tilde{U}_{\mu}(n)$ associated to improved links are not necessarily part of the gauge group anymore. To reproject to the original group we find the $\Sp{2N}$ matrix $U_{\mu}(i)$ that maximizes $\re \tr \tilde{U}^{\dag} U$: first, a crude projection is operated on $\tilde{U}^{N}$ using a resymplecticisation algorithm, see~\cite{thiscontrib194} for our choice, second, a certain number of cooling~\cite{Hoek:1986nd} steps (we used 15) is performed on the link. This allows to iteratively approach the sought for $U$. 

Starting from $M$ elementary basis paths in a given symmetry channel, $N$ iterations of the improvement process results in a collection $N\times M$ operators. Denoting with $\left\{\phi_i(t)\right\}$ this collection in a given symmetry channel, we may compute the correlation matrix 
\begin{equation}\label{eq:glue-corr-2}
	\Gamma_{ij}(t) = \frac{ \langle 0 | \phi_i(t) \phi_j^\dag(0)| 0 \rangle }{  \langle 0 | \phi_i(0) \phi_j^\dag(0)| 0 \rangle }
\end{equation}
and diagonalize it, assuming maximal rank. If $v_i$ is the eigenvector corresponding to the greatest eigenvalue, the operator
\begin{equation}
	\Phi(t) = \sum_{i=0}^{N\times M} v_i \phi_i(t)
\end{equation}
creates the state of maximum overlap with the group state of the given symmetry channel. The corresponding mass can then be extracted from a fit of
\begin{equation}\label{eq:masscorrfinal}
\tilde{\Gamma}_{ii}(t) = A_i \cosh{\left( m_i t- \frac{N_t}{2}\right)}
\end{equation}
to the data, where $\tilde{\Gamma}$ is diagonalized correlation matrix, or by looking for a plateau value in
\begin{equation}
	m_\text{eff}(t) = -\log{  \frac{  \tilde{\Gamma}_{ii}(t) }{ \tilde{\Gamma}_{ii}(t-1) }  }  
\end{equation}
Note that in both cases, the extracted mass also receives contributions from excited states in the same symmetry channel.

In the case of fluxtube states, we are interested in extracting the string tension, i.e. the proportionality constant between the energy and the length quantum number
\begin{equation}
	\sigma = \lim_{L\to\infty} \frac{m(L)}{L}~.
\end{equation}
Note that the corresponding correlator may be measured in several channels, depending on our (entirely conventional) choice of what direction is ``time-like''. However, since we are in a euclidean setup, we expect the final results for the string tensions $\sigma_s$ and $\sigma_t$ to be compatible. Assuming that the string in question is bosonic in nature, the corrections to the formula above for finite $L$ contribute with a power series in $1/L$ and can be computed in the framework of Effective String Theory, see~\cite{Aharony:2013ipa} and references therein. There it is shown that the request of Poincar\'e invariance of the effective action strongly constrains the form of the power series coefficients. As a result, all the correction terms in $D=4$ are universal up to order $1/L^5$ and coincide with a Taylor expansion of the Nambu Goto (NG) spectrum in $1/L$ around $L=\infty$. In particular we have, ,
\begin{equation}
	m_l^{LO}(L) = \sigma L  - \frac{\pi (D-2) }{6 L},\quad m_l^{NLO}(L)  = \sigma  L - \frac{\pi (D - 2)}{6 L} + \frac{1}{2}\left( \frac{\pi (D - 2)}{6}\right)^2 \frac{1}{\sigma L^2}~.
\end{equation}
at the Leading (LO) and Next-to-leading (NLO) order, and
%\begin{equation}\label{eq:effs_NLO}
%	m_l^{NLO}(L)  = \sigma  L - \frac{\pi (D - 2)}{6 L} + \frac{1}{2}\left( \frac{\pi (D - 2)}{6}\right)^2 \frac{1}{\sigma L^2} \ . 
%\end{equation}
\begin{equation}\label{eq:effs_NG}
	m_l^{NG}(L) = \sigma L \sqrt{1 - \frac{(D-2) \pi}{3\sigma L^2}}~.
\end{equation}
for the complete NG spectrum. Estimates of the string tension at these orders of approximation were obtained by inverting the above formulas for $\sigma$.

\section{Numerical Setup and results}

The simulations were performed using a version of the HiRep code suite approriately modified to accomodate the $\Sp{2N}$ series of gauge theories. In particular, we work at $N=2$. For the variational procedure with improved operators, we employed the same automatized algorithm used in~\cite{Lucini:2010nv}. Operators were blocked to the level $N_b\leq L$ with smearing and blocking parameters chosen as $(p_a=0.4,~p_b=0.16)$. After $N_c=15$ steps of cooling, This resulted in a variational basis of $\sim 200$ operators. For each value of the lattice volume and of the coupling that were probed, $10000$ configurations were stored to be later analyzed.

\begin{figure}[!th] % no figure before 1st section
	\small
  \centering
  \sidecaption
  \includegraphics[width=0.7\textwidth,clip]{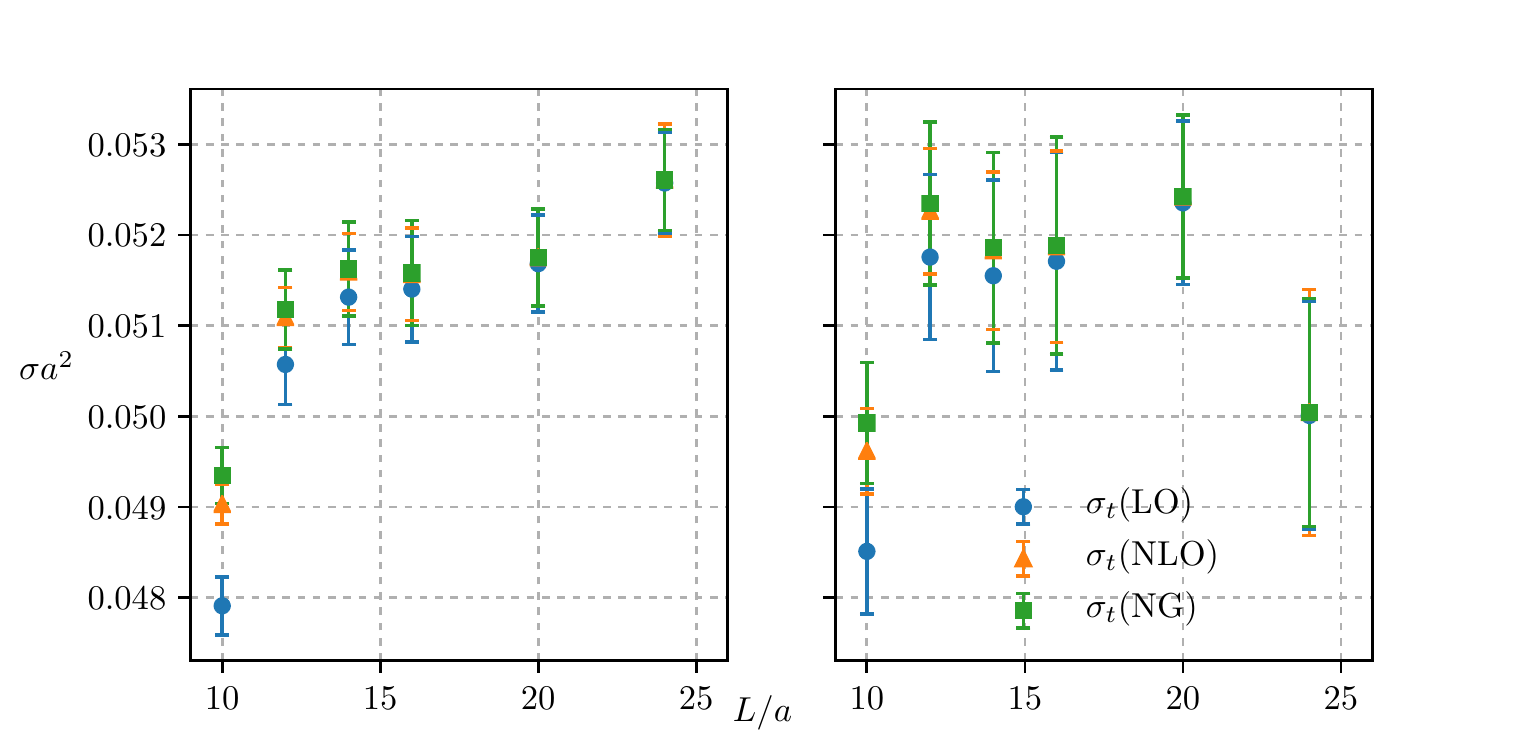}
  \caption{Estimates of the string tension in the spacelike ($\sigma_s$) and timelike ($\sigma_t$) channels at various orders of approximation and at various values of the lattice volume for $\beta=7.7$. The systematic effect on the largest lattices is explained by the difficulty in extracting a very large mass from the correlator eq.~(\ref{eq:masscorrfinal}).}
  \label{fig:stension}% Give a unique label
\end{figure}

\begin{figure}[!thb] % no figure before 1st section
	\begin{subfigure}{0.5\textwidth}
	\centering
  	\includegraphics[width=0.9\textwidth,clip]{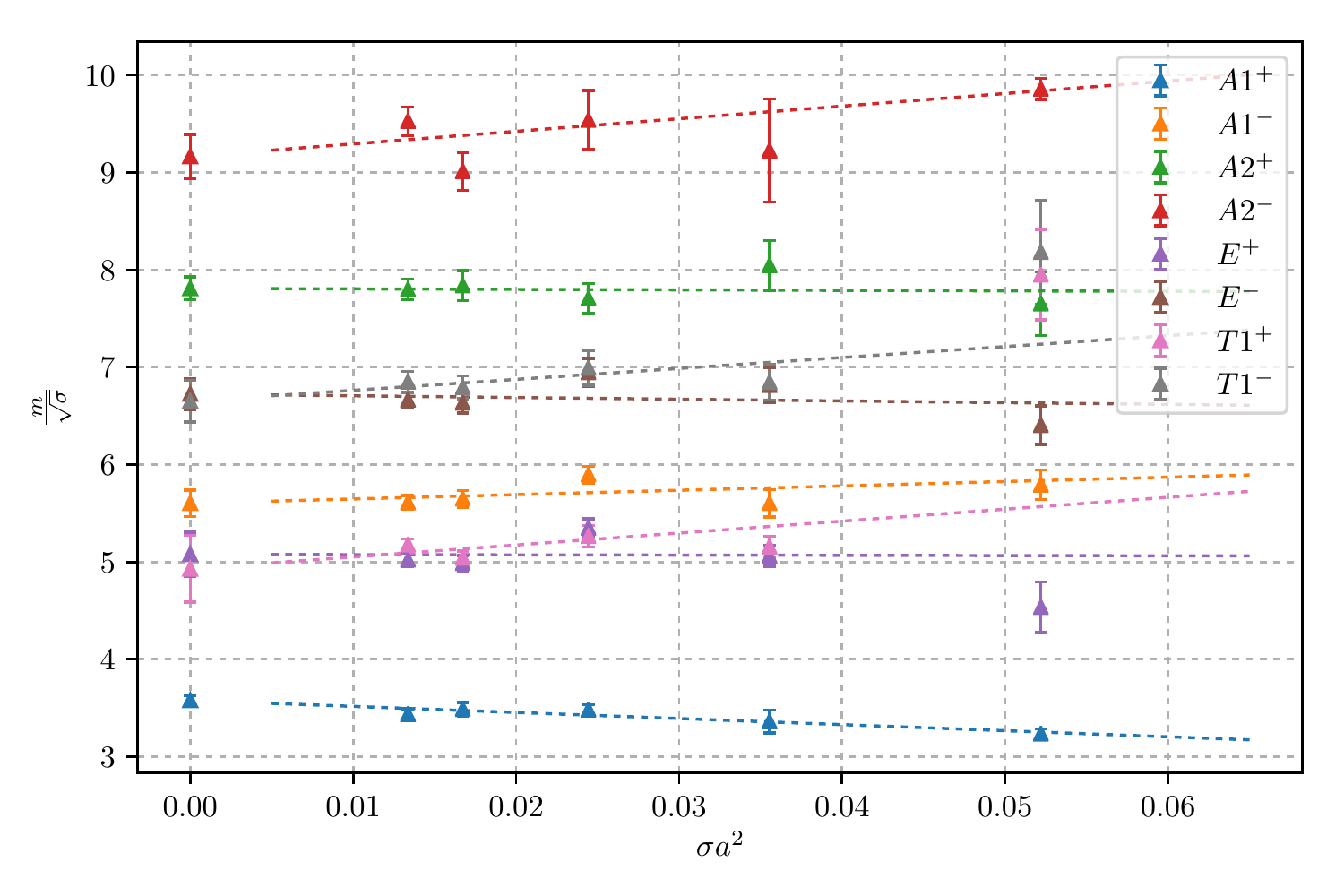}
	\end{subfigure}
	\begin{subfigure}{0.5\textwidth} % no figure before 1st section
  	\centering
  	\includegraphics[width=0.9\textwidth,clip]{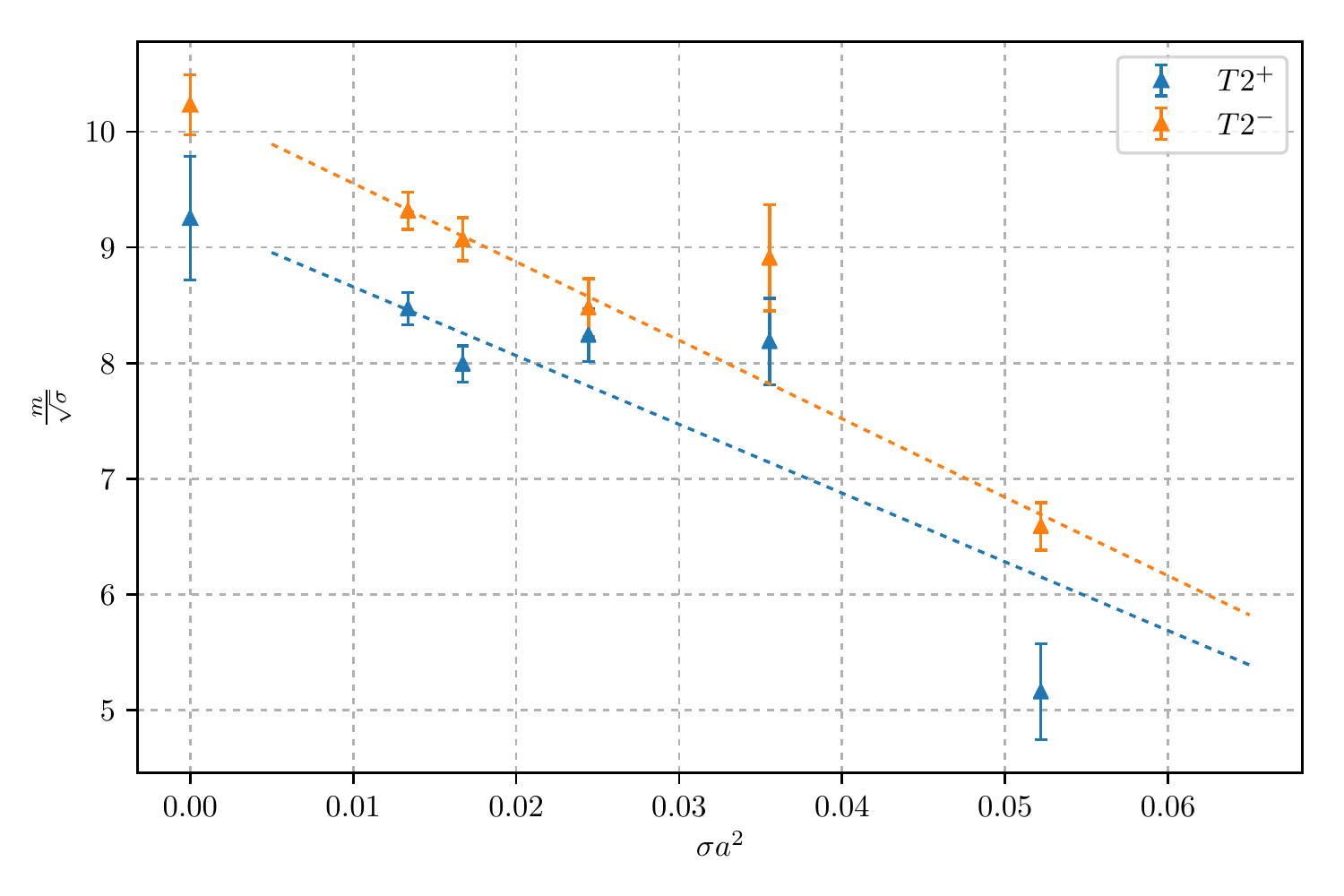}
	\end{subfigure}
	\caption{Continuum limit extrapolations for glueball masses. For each symmetry channel $R^P$, we expect the ratio $m_G/\sqrt{\sigma}$ to tend to its continuum limit value according to eq.~\ref{eq:contlim}. Note, in addition, that the estimates in the channels $E^+, T1^+$ and $E^-, T1^-$, respectively, tend to the same value as $a\to 0$, as we expect if Poincar\'e invariance is to be restored in the continuum limit.}
	\label{fig:extrap}
\end{figure}

Data was initially collected on a $(La = 10 a)^4$ lattice at $\beta=7.7$. This choice of the coupling guarantees that we are in the confined phase. After estimates of the string tension and the glueball masses were computed, the process was repeated for larger lattice sizes at the same coupling. This allowed to estimate the importance of finite size effects. Additionally, we explicitly checked that the estimates of the string tension obtained at each lattice volume, measured in different channels, were statistically compatible, see the right hand side of tab.~\ref{tab:sigmas1} and fig.~\ref{fig:stension}. Our final estimate thereof was then computed as the weighted average of the two results, using the inverse errors as weights. The final estimates of the string tension at finite lattice spacings are reported in tab.~\ref{tab:sigmas1}. Results for the glueball masses at finite lattice spacing are reported in tab.~\ref{tab:gluevol}. Once a lattice large enough for the finite size effects to be neglectable was found, we used the scale setting results reported in~\cite{thiscontrib194} to shrink the lattice spacing at (approximately) constant physical volume. Computing the ratio $m_G/\sqrt{\sigma}$ for each of the couplings and lattice volumes shown in the first two columns of the left hand side of tab.~\ref{tab:sigmas1}, and assuming discretization effects to behave as
\begin{equation}\label{eq:contlim}
	\frac{m_G}{\sqrt{\sigma}} ( a ) - \frac{m_G}{\sqrt{\sigma}} ( 0 ) \propto \sigma a^2
\end{equation}
we could extrapolate the above ratio to the continuum limit for each symmetry channel. The extrapolations are visible in fig.\ref{fig:extrap} and the continuum limits reported in tab.~\ref{tab:contlim}. Note that the estimates for the $E^\pm$ and the $T1^\pm$ channels are numerically compatible, as we expect if Poincar\'e symmetry is to be restored in the continuum limit.

\section{Conclusions}

In this work we have carried out the first numerical estimation of the glueball spectrum in the pure $\Sp{4}$ gauge theory. After adapting the HiRep code suite to simulate the $\Sp{2N}$ gauge theory, we employed a fully automatised process to determine the best-overlap variational basis built from improved operators. This allowed us to overcome the exponentially-decaying signal-to-noise ratio problem and to obtain precision measurements of glueball masses in units of the square root of the string tension in all the symmetry channels. The magnitude of our results is comparable to what is found in $\SU{N}$ theories, see ~\cite{Chen:2005mg, Lucini:2004my} and reference therein, and the value of the 
\begin{equation}
\eta=\frac{m^2(0^{+})}{\sigma}\cdot \frac{C_2(F)}{C_2(A)}
\end{equation}
ratio, $\eta(Sp(4))= 5.27(15) $ is compatible with the expectations from~\cite{Hong:2017suj}. 

\begin{table}[!ht]
	\small
	\centering
	%\sidecaption
	\begin{tabular}{|c||c|c||c|}
	\hline
	$L/a$ & $\sqrt{\sigma_t} a$(NG) & $\sqrt{\sigma_s} a$(NG) & $\sqrt{\sigma} a$\\
	\hline
	$ 24 $ &  $ 0.2294(12) $ & $ 0.2237(28) $ & $ 0.2285(11) $ \\
	$ 20 $ &  $ 0.2275(12) $ & $ 0.2290(20) $ & $ 0.228(10) $ \\
	$ 16 $ &  $ 0.2271(13) $ & $ 0.2278(26) $ & $ 0.2272(11) $ \\
	$ 14 $ &  $ 0.2272(11) $ & $ 0.2277(23) $ & $ 0.2273(10) $ \\
	$ 12 $ &  $ 0.22623(96) $ & $ 0.2288(20) $ & $ 0.22673(86) $ \\
	$ 10 $ &  $ 0.22215(69) $ & $ 0.2234(15) $ & $ 0.22238(63) $ \\
	\hline	
	\end{tabular}
	\qquad
	\begin{tabular}{|c|c||c|c||c|}
	\hline
	$L/a$ & $\beta$ & $\sqrt{\sigma} a$\\
	\hline
	$32$ & $8.3$ & $0.1156(3)$ \\
	$26$ & $8.2$ & $0.1293(6)$ \\
	$20$ & $8.0$ & $0.1563(6)$ \\
	$18$ & $7.85$ & $0.1885(7)$\\
	$16$ & $7.7$ & $0.227(1)$\\
	\hline	
	\end{tabular}
	\caption{On the left hand side, in the first two columns, the value of $\sqrt{\sigma}$ extracted from the full lightcone spectrum (NG) prediction for the ground state energy of the fluxtube length $L$ at $\beta=7.7$. In the last column, weighted averages over the timelike and spacelike channels. On the right hand side, the final estimates for $\sqrt{\sigma}$ at different lattice setups.}
	\label{tab:sigmas1}
\end{table}
%\begin{table}
%	\small
%	\centering
%	\begin{tabular}{|c||c|c||c|}
%	\hline
%	$L/a$ & $\sqrt{\sigma_t} a$(NG) & $\sqrt{\sigma_s} a$(NG) & $\sqrt{\sigma} a$\\
%	\hline
%	$ 24 $ &  $ 0.2294(12) $ & $ 0.2237(28) $ & $ 0.2285(11) $ \\
%	$ 20 $ &  $ 0.2275(12) $ & $ 0.2290(20) $ & $ 0.228(10) $ \\
%	$ 16 $ &  $ 0.2271(13) $ & $ 0.2278(26) $ & $ 0.2272(11) $ \\
%	$ 14 $ &  $ 0.2272(11) $ & $ 0.2277(23) $ & $ 0.2273(10) $ \\
%	$ 12 $ &  $ 0.22623(96) $ & $ 0.2288(20) $ & $ 0.22673(86) $ \\
%	$ 10 $ &  $ 0.22215(69) $ & $ 0.2234(15) $ & $ 0.22238(63) $ \\
%	\hline	
%	\end{tabular}
%	\caption{In the first two columns, the value of $\sqrt{\sigma}$ extracted from the full lightcone spectrum (NG) prediction for the ground state energy of the fluxtube length $L$ at $\beta=7.7$. In the last column, the result of eq.~(). }
%	\label{tab:sigma_7.7}
%\end{table}
%
%\begin{table}
%	\small
%	\centering
%	\begin{tabular}{|c|c||c|c||c|}
%	\hline
%	$L/a$ & $\beta$ & $\sqrt{\sigma} a$\\
%	\hline
%	$32$ & $8.3$ & $0.1156(3)$ \\
%	$26$ & $8.2$ & $0.1293(6)$ \\
%	$20$ & $8.0$ & $0.1563(6)$ \\
%	$18$ & $7.85$ & $0.1885(7)$\\
%	$16$ & $7.7$ & $0.227(1)$\\
%	\hline	
%	\end{tabular}
%	\caption{The final estimates for $\sqrt{\sigma}$ at different lattice setups.}
%	\label{tab:sigmas}
%\end{table}

\begin{table}[!ht]
	\small
	\centering
	\begin{tabular}[!htb]{ |c||c|c|c|c|c|c| }
	\hline
	$R^P$ & $L=10$ & $L=12$ & $L=14$ & $L=16$ & $L=20$ & $L=24$ \\
	\hline
	$ A1^+ $ &   $0.569(13)$ &  $0.728(15)$ &  $0.738(16)$ &  $0.742(16)$ &  $0.764(15)$ &  $0.739(11)$ \\ 
	$ A1^- $ &   $1.039(35)$ &  $1.275(41)$ &  $1.406(47)$ &  $1.210(40)$ &  $1.300(34)$ &  $1.323(34)$ \\ 
	$ A2^+ $ &   $1.70(11)$ &  $1.706(95)$ &  $1.778(95)$ &  $1.650(76)$ &  $1.771(81)$ &  $1.748(74)$ \\ 
	$ A2^- $ &   $2.48(34)$ &  $1.83(17)$ &  $1.74(14)$ &  $2.21(25)$ &  $2.23(24)$ &  $2.252(22)$ \\ 
	$ E^+ $ &   $0.623(13)$ &  $1.111(32)$ &  $1.150(32)$ &  $1.159(27)$ &  $1.217(26)$ &  $1.036(59)$ \\ 
	$ E^- $ &   $1.402(62)$ &  $1.347(58)$ &  $1.401(48)$ &  $1.509(66)$ &  $1.597(59)$ &  $1.463(45)$ \\ 
	$ T1^+ $ &   $1.170(43)$ &  $1.220(36)$ &  $1.202(39)$ &  $1.209(31)$ &  $1.173(26)$ &  $1.82(11)$ \\ 
	$ T1^- $ &   $1.465(75)$ &  $1.513(69)$ &  $1.515(66)$ &  $1.522(57)$ &  $1.499(51)$ &  $1.87(12)$ \\ 
	$ T2^+ $ &   $1.53(11)$ &  $1.70(12)$ &  $1.99(14)$ &  $1.578(94)$ &  $1.839(96)$ &  $1.179(95)$ \\ 
	$ T2^- $ &   $1.60(12)$ &  $2.04(18)$ &  $2.38(27)$ &  $2.07(18)$ &  $1.94(15)$ &  $1.505(46)$ \\ 
	$ m_s $ &   $0.3744(32)$ &  $0.5196(53)$ &  $0.6436(73)$ &  $0.7570(93)$ &  $0.981(11)$ &  $1.218(13)$ \\ 
	$ m_t $ &   $0.3804(69)$ &  $0.534(11)$ &  $0.647(15)$ &  $0.762(19)$ &  $0.995(18)$ &  $1.157(30)$ \\ 
	\hline
	\end{tabular}
	\caption{Estimates of the ratio $m_G/\sqrt{\sigma}$ for each symmetry channel $R^P$, obtained by fitting eq.~\ref{eq:masscorrfinal} to the data at coupling $\beta=7.7$. The operators considered were blocked at level $N_b\leq L$, with $N_c=15$ cooling steps.}
	\label{tab:gluevol}
\end{table}

\begin{table}[!tb]
	\small
	\centering
	\begin{tabular}{|c|c|}
	\hline
	$R^P$ & $m(R^P)/\sqrt{\sigma}$\\
	\hline
	\input{cont2}
%	\hline
%	\end{tabular}
%	\qquad
%	\begin{tabular}{|c|c|}
%	\hline
%	$R^P$ & $m(R^P)/\sqrt{\sigma}$\\
%	\hline
	\input{cont1}
	\hline
	\end{tabular}
	\caption{Continuum limit estimates of glueball masses in all the symmetry channels. These values are the intercepts for the straight lines in fig.~\ref{fig:extrap} for each one of the symmetry channels $R^P$ considered.}
	\label{tab:contlim}
\end{table}

\clearpage

\bibliography{lattice2017}
%
%%%%%%%%%%%%%%%%%%%%%%%%%%%%%%%%%%%%%%%%%%%%%%%%%%%%%%%%%%%%%%%%%%%%%%%%%%%%%%
\end{document}

%% file: cont2.tex
$ A1^+ $ & $ 3.557(52) $ \\
%$ A1^{+*} $ & $ 6.05(4) $ \\
$ A1^- $ & $ 5.74(19) $ \\
%$ A1^{-*} $ & $ 7.81(8) $ \\
$ A2^+ $ & $ 7.91(17) $ \\
$ A2^- $ & $ 9.42(40) $ \\

%% file: cont1.tex
$ E^+ $ & $ 5.02(16) $ \\
$ E^- $ & $ 6.61(13) $ \\
$ T1^+ $ & $ 5.070(91) $ \\
$ T1^- $ & $ 6.872(89) $ \\
$ T2^+ $ & $ 8.73(30) $ \\
$ T2^- $ & $ 9.50(37) $ \\